# Low-temperature Synthesis of FeTe$_{0.5}$Se$_{0.5}$ Polycrystals with a High Transport Critical Current Density

Qing-Ping Ding[1,2], Shyam Mohan[1], Yuji Tsuchiya[1], Toshihiro Taen[1], Yasuyuki Nakajima[1,2] and Tsuyoshi Tamegai[1,2]

[1]Department of Applied Physics, The University of Tokyo, 7-3-1 Hongo, Bunkyo-ku, Tokyo 113-8656, Japan
[2]JST, Transformative Research-Project on Iron Pnictides (TRIP), 7-3-1 Hongo, Bunkyo-ku, Tokyo 113-8656, Japan

**Abstract.** We have prepared high-quality polycrystalline FeTe$_{0.5}$Se$_{0.5}$ at temperature as low as 550 ℃. The transport critical current density evaluated by the current-voltage characteristics is over 700 A/cm$^2$ at 4.2 K under zero field, which is several times larger than FeTe$_{0.5}$Se$_{0.5}$ superconducting wires. The critical current density estimated from magneto-optical images of flux penetration is also similar to this value. The upper critical field of the polycrystalline FeTe$_{0.5}$Se$_{0.5}$ at $T = 0$ K estimated by Werthamer-Helfand-Hohenberg theory is 585 kOe, which is comparable to that of single crystals. This study gives some insight into how to improve the performance of FeTe$_{0.5}$Se$_{0.5}$ superconducting wires.

Motivated by the discovery of superconductivity in oxypnictide LaFeAsO$_{1-x}$F$_x$ with a critical temperature $T_c \sim 26$ K, a series of iron-based superconductors have been synthesized in a short time [1-8]. Among these compounds, FeSe and FeTe$_{1-x}$Se$_x$ (11 system) attracted much attention because of its simpler planar crystal structure than layered iron-pnictides. Therefore, they are better candidates to investigate the underlying mechanism of superconductivity in the iron-based system. The lower toxicity and less sensitivity to atmosphere of starting materials also make 11 system more promising compared with iron-pnictide superconductors. Although the transition temperature, $T_c$, is not very high in this 11 system, it can be enhanced to as high as 37 K by applying pressure [9, 10]. Recently, by introducing alkali metal K (Cs, Rb) into this 11 system to form a ternary compound, $T_c$ is increased up to 30 K [11-13]. In addition to the very simple crystal structure, the very high values of upper critical field make this system a potential candidate for practical applications [14-18]. Also critical current density, $J_c$, in a high-quality single crystal is reported to be over $1 \times 10^5$ A/cm$^2$ [18-20]. Superconducting wires made from FeTe$_{0.5}$Se$_{0.5}$ have been reported, but the transport critical current density is restricted to ~ 230 A/cm$^2$ at 4.2 K even under zero field [21-23]. Polycrystalline samples of iron-based superconductors in general have the problem of weak links. In order to improve the performance of FeTe$_{1-x}$Se$_x$ wires, it is essential to investigate the performance of FeTe$_{1-x}$Se$_x$ polycrystals and how to reduce the weak-link features. We have reported FeTe$_{1-x}$Se$_x$ polycrystals prepared at higher temperatures, 600 ℃ ~ 680 ℃, with less impurities [24]. Since the grain sizes prepared at higher temperatures are too large, weak links between grains are particularly obvious. Here we report the low-temperature synthesis and characterization of high quality FeTe$_{1-x}$Se$_x$ polycrystalline sample. The characterization through X-ray diffraction, magnetization, resistivity, transport critical current density, and magneto-optical measurements are discussed.

FeTe$_{0.5}$Se$_{0.5}$ polycrystalline samples were synthesized by solid-state reaction method. Stoichiometric amounts of Fe powder (Kojundo Chemical Laboratory, 99.9%), Te grains (same as above, 99.9999%), and Se grains (same as above, 99.9%) were weighed, mixed and ground with an agate mortar and pestle. The mixing and grinding processes were carried out in nitrogen-filled glove box. The ground powder was cold-pressed into pellets with 600 kg/cm$^2$ uniaxial pressure. The pellets were sealed in an evacuated quartz tube, and slowly ramped to 550 ℃ in 5 hours from room temperature, and kept for 24 hours. After cooling to room temperature naturally, the reacted pellets were reground, pressed (1800 kg/cm$^2$ uniaxial pressure), sealed in a quartz tube, sintered again at 550 ℃ for 24 h, and finally annealed at 400 ℃ for 24 h, followed by quenching. When the sintering temperature is lower than 500 ℃ and higher than 400 ℃, although 11 phase can be formed, there are more impurities, and the superconducting volume fraction is very small.

The phase identification of the sample was carried out by means of powder X-ray diffraction (M18XHF, MAC Science) with Cu-$K\alpha$ radiation generated at 40kV and 200 mA. Bulk magnetization is measured by a superconducting quantum interference device SQUID magnetometer (MPMS-5XL, Quantum Design). The microstructures were characterized by Scanning Electron Microscopy (SEM, Hitachi S-4300) operated at 15 kV. Resistivity and current-voltage (*I-V*) measurements were performed by four-probe method with silver paste for contact. Resistivity measurements were performed within the sample chamber of a SQUID magnetometer. *I-V* measurements were performed in a bath type cryostat (Spectromag, Oxford Instruments). For local magnetic characterization, we applied magneto-optical (MO) imaging. For this purpose, the sample was cut using a wire saw, and the surface was polished with a lapping film. A Bi-substituted iron-garnet indicator film is placed in direct contact with the sample, and the whole assembly is attached to the cold finger of a He-flow cryostat (Microstat-HR, Oxford Instruments) and cooled down to 5 K. MO images are acquired by using a cooled CCD camera with 12-bit resolution (ORCA-ER, Hamamatsu). To enhance the visibility of the local magnetic induction and eliminate the signals from the impurity phases, a differential imaging technique is employed [25, 26].

Figure 1(a) shows the X-ray diffraction pattern of as-prepared polycrystalline sample. All the peaks were well indexed using a space group of P4/nmm except for small peaks as an impurity phase marked by asterisks. The compound crystallizes in a tetragonal structure and the impurity phase is identified as hexagonal $Fe_7Se_8$. The calculated lattice constants are $a =$ 0.3799 nm, and $c =$ 0.5980 nm, which are slightly smaller than the samples prepared at higher temperatures ($a =$ 0.3800 nm, and $c =$ 0.6011 nm) and single crystals [19, 27]. Figure 1(b) is the SEM image of this sample. From this image, a typical grain size is ~ 10 μm or less although with a broad size distribution. The grains are sheet-like with the ratio of width (*w*) to thickness (*t*) may be larger than 10. Although the amount of impurities of this sample is higher than the sample prepared at commonly used higher temperature at 600-680 ℃ [24], the average grain size has been reduced from 50 ~ 100 μm to less than 10 μm. Temperature dependences of zero-field-cooled (ZFC) and field-cooled (FC) magnetization at 5 Oe of the $FeTe_{1-x}Se_x$ polycrystalline sample are shown in Fig. 1(c). The sample shows an onset of diamagnetism at around 12.5 K. The feature of two transitions, which is due to weak links between grain boundaries and appeared in many $FeTe_{1-x}Se_x$ polycrystalline samples prepared at higher temperatures [22, 24], is suppressed in our present sample. So we think the reduction of grain size is beneficial to the grain boundaries.

Figure 2(a) shows temperature dependence of the resistivity of the sample. The room temperature resistivity is 690 μΩ cm. The resistivity shows semiconducting behavior above 150 K, and shows metallic behavior in the normal state below 150 K. A sharp drop in resistivity was observed at about 14.3 K, which indicates the onset of superconductivity. The zero resistance occurs at 12.5 K and the transition width is 1.8 K. The residual resistivity ratio $\rho(300 \text{ K})/\rho(T_c^{onset})$ is 2.06. Similar temperature dependence of resistivity is reported for samples with compositions close to $FeTe_{0.5}Se_{0.5}$ [19, 22, 27-31]. This behavior has been attributed to weak charge-carrier localization due to a large amount of excess Fe in $Fe_{1+y}Te_{1-x}Se_x$ system [27]. The value of resistivity in our polycrystalline sample is larger than Ref [27], but similar to Ref [19], and lower than that reported in Refs [22, 28, 29]. The intrinsic microscopic inhomogeneities of polycrystalline sample and impurity phase may be the reason for this higher value of resistivity. The variation of $T_c$ with magnetic field is shown in figure 2 (b) for $H$ = 0, 10, 20, 30, 40 and 50 kOe. With increasing field, the resistive transition shifts to lower temperatures accompanied by a slight increase in the transition width. Inset of figure 2 (b) shows the variation of the upper critical field $H_{c2}$ with reduced temperature $t = T/T_c^{onset}$ for the FeTe$_{0.5}$Se$_{0.5}$ polycrystalline sample. The values of $H_{c2}$ were defined as the field at the midpoint of the resistive transition. The slope of $H_{c2}$ at $T_c$ is -60.6 kOe K$^{-1}$. The value of $H_{c2}$ at $T$ = 0 K estimated using the Werthamer–Helfand–Hohenberg formula [32], $H_{c2}(0)=−0.69T_c|dH_{c2}/dT|_{T=Tc}$, is 585 kOe. This value is comparable to the single crystals [14-18]. In order to extract the superconducting parameters, we have used the Ginzburg–Landau (GL) formula for the coherence length ($\xi$), $\xi= (\Phi_0/2\pi H_{c2})^{1/2}$, where $\Phi_0$ = $2.07 \times 10^{-7}$ G cm$^2$, the coherence length $\xi$ at the zero temperature is calculated as 2.38 nm. It is well known that, for the two-gap superconductivity in the dirty limit, the impurity scattering could strongly enhance the upper critical field $H_{c2}$. This phenomenon has been clearly observed in MgB$_2$, a dirty two-gap superconductor, where the $H_{c2}$ increases remarkably by alloying MgB$_2$ with some impurities [33]. Therefore, the very high $H_{c2}$ value observed in our polycrystals could be reasonably attributed to the strong impurity scattering effect from the impurities in such a new mutiband superconductor.

Figure 3(a) shows magnetic hysteresis curves of this sample. A small amount of Fe$_7$Se$_8$ impurities may be the reason for the ferromagnetic background in *M-H* curves. Shown in the inset is *M-H* curve at 15 K, which is higher than $T_c$. Since the curves at 12.5 K (not shown) and 15 K are almost identical, we assumed this background is temperature-independent at low temperatures. In these hysteresis loops, the ratio between superconducting signals to the ferromagnetic background is not as strong as that of the sample prepared at higher temperatures [24]. This is because the magnetization is proportional to the grain sizes if the intragranular critical current density ($J_c^{intra}$) is fixed, and the grains prepared at higher temperatures are much larger. From the magnetization hysteresis loops, we can evaluate intragranular critical current density $J_c^{intra}$ for polycrystalline samples using the Bean's model with the assumption of field–independent $J_c$. According to Bean's model, $J_c^{intra}$ [A/cm$^2$] is given by

$$J_c^{intra} = 30\frac{\Delta M}{d}, \qquad (1)$$

with an assumption that intergranular critical current is zero, where $\Delta M$[emu/cc] is $M_{down}$ - $M_{up}$, $M_{up}$ and $M_{down}$ are the magnetization when sweeping field up and down, respectively, $d$[cm] is the average diameter of the grain in the polycrystalline sample [34]. The $\Delta M$ used here is after subtracting the ferromagnetic background. Figure 3(b) shows the field dependences of the obtained $J_c$ for this sample calculated from the data shown in figure 3(a) using Eq. (1) and the typical dimension. $J_c$ calculated from *M-H* curve is estimated to be $2 \times 10^5$ A/cm$^2$ at 5 K under zero field, this value is a little smaller than that in the single crystal [19], but still in the range for applications. $J_c$ values in excess of $5 \times 10^4$ A/cm$^2$ are sustained up to 10 kOe.

Figures 4(b)-(e) depict MO images of the FeTe$_{0.5}$Se$_{0.5}$ polycrystalline sample in the remanent state after applying a 500 Oe field for 0.25 seconds which subsequently reduced down to zero. Figure 4(a) shows the polished surface of the FeTe$_{1-x}$Se$_x$ polycrystalline sample for MO measurements. The sample dimensions are 895×555×200 μm$^3$. Shown in figures 4(b) to 4(e) are MO images of the remanent state recorded at several different temperatures. These images are similar with the MO images of 1111 polycrytals [35-37]. In these figures, the bright regions correspond to the trapped flux in the sample. The dots in these images are due to defects in the indicator garnet film. At all temperatures, the field profile is inhomogeneous, which implies that the intergranular current density is much smaller compared with the intragranular current density. The intragranular current density decreases gradually as the temperature is increased towards $T_c$. We calculated the intragranular critical current density from the magnetic induction profile. Figure 4(f) shows the magnetic induction profiles along the dotted line in figure 4(b). In this calculation, we roughly estimate the intragranular critical

current densities by $J_c \sim dB/dx$. For typical grains, $J_c$ thus estimated is $\sim 1 \times 10^4$ A/cm$^2$ at 5 K. This value is much smaller than that estimated from the *M-H* curve as shown in figure 3(b). This is because the formula we adopted here is based on the condition with infinite thickness of the sample, but from the SEM image, the ratio of width to thickness is very large. So the real $J_c$ should be obtained by multiplying $w/t$, which is roughly 10.

In figure 3 and figure 4, both *M-H* and MO images of the remanent state for FeTe$_{1-x}$Se$_x$ polycrystalline sample only give the intragranular $J_c$. For practical applications and understanding the superconducting mechanism, global critical current density $J_c$ should be investigated. In iron-based superconductors, global $J_c$ for polycrystalline samples were less than 10$^4$ A/cm$^2$, and very sensitive to magnetic field [38, 39]. Namely, weak-link feature is common in these systems. This has been usually ascribed to the tensile strain generated by dislocations located along the boundary [40] or competing orders, low carrier density and unconventional pairing symmetry [41]. We investigated the transport $J_c$ of FeTe$_{0.5}$Se$_{0.5}$ polycrystalline sample. The sample dimension for this measurement is 1200×180×75 μm$^3$, and the distance between two voltage contacts is 470 μm. In superconductors, driving force competes with pinning force. If the former wins over the latter, measurable voltages could be observed. Inset of figure 5(a) shows the zero-field *I –V* characteristics at different temperatures ranging from 4.2 K to 10 K. Here we adopt $E$ = 1 μV/cm as a criterion for the *I –V* curve to define transport $J_c$. Figure 5(a) shows the temperature dependence of transport $J_c$ at zero field. At 4.2 K, a transport $J_c$ as high as 724 A/cm$^2$ was observed in the FeTe$_{1-x}$Se$_x$ polycrystalline sample. Transport $J_c$ does not increase so fast below $T_c$. Transport $J_c$ as a function of field at 4.2 K is shown in figure 5(b). Similar to YBa$_2$Cu$_3$O$_{7-y}$ and other iron-based superconductors, transport $J_c$ shows strong field dependence at low fields. For example, $J_c$ (0 Oe) is 10 times as large as $J_c$ (0.5 kOe). When the field is more than 0.5 kOe, the field dependence of $J_c$ is not as strong as that in the low field region. The value of transport $J_c$ is more than two orders less than the intragranular $J_c$ as shown in the figure 3(b) and that obtained from the magnetic induction profile in figure 4(e). This implies the presence of weak links between superconducting grains. The transport $J_c$ of FeTe$_{1-x}$Se$_x$ polycrystalline sample is smaller than SmFeAsO$_{0.7}$F$_{0.3-\delta}$ wires (1300 A/cm$^2$) [38] and Ag doped Sr$_{0.6}$K$_{0.4}$Fe$_2$As$_2$ wires (3750 A/cm$^2$) [39], but the same order as randomly oriented polycrytalline YBa$_2$Cu$_3$O$_{7-y}$ [42], but several times larger than that in FeTe$_{0.5}$Se$_{0.5}$ wires [21-23].

Figures 6(a) to 6(d) reveal the penetration of vortices at 5 K in the FeTe$_{0.5}$Se$_{0.5}$ polycrystalline sample. In figure 6(a), at low field of 2 Oe, most part of the sample was still in the Meissner state. When the field was increased to 10 Oe, the flux profile shows a pattern similar to the critical state as shown in figure 6(c). When the field was further increased to 20 Oe, the sample was fully penetrated. For samples prepared at higher temperature [21], when the field was as low as 1 Oe, magnetic flux penetrates intergranular regions. The shapes of the superconducting grains are more visible when the field is increased. This also indicated the weak-link feature in this lower temperature prepared sample is much reduced compared with the higher temperature prepared sample [21]. The intergranular $J_c$ for a thin superconductor can be roughly estimated by $J_c \sim \Delta B/t$, where $\Delta B$ is the difference in local induction between the centre and edge of the sample and $t$ is the thickness of the sample. With $\Delta B \sim 10$ G and $t$ = 200 μm, $J_c$ is estimated to be 500 A/cm$^2$ at 5 K, which is consistent with the value 440 A/cm$^2$ measured from transport measurement (figure 5(a)).

In superconducting wires, polycrystalline samples with smaller grain size are more promising. Our results show that both grain size and weak-link feature has been reduced in the polycrystalline FeTe$_{0.5}$Se$_{0.5}$ samples prepared at 550 ℃ compared with the samples prepared at higher temperatures. By using higher static pressure to make denser pellets and increasing the annealing time, or using high-pressure synthesis method we can obtain more pure and homogenous samples with higher transport critical current density. Alternatively, we can utilize the powder-in-tube method to make the sample more dense and well-connected. This is the subject of ongoing research.

In conclusion, X-ray diffraction, magnetization, resistivity, transport critical current density and magneto-optical measurements were performed on high quality polycrystalline FeTe$_{0.5}$Se$_{0.5}$ prepared at 550 °C. The transport critical current density over 700 A/cm$^2$ at 4.2 K under zero field is obtained. This value is several times larger than that of FeTe$_{0.5}$Se$_{0.5}$ superconducting wires, although it is more than two orders less than the intragranular critical current density. $J_c$ estimated from magneto-optical (MO) images of flux penetration is also similar to this value. The upper critical field $H_{c2}$ of the polycrystalline FeTe$_{0.5}$Se$_{0.5}$ is 585 kOe, which is comparable to that of single crystals. Our low-temperature synthesis is promising to the development of practical FeTe$_{1-x}$Se$_x$ system, with potential higher transport critical current density, for applications especially for superconducting wires.

**References**

[1] Kamihara Y, Watanabe T, Hirano M and Hosono H 2008 *J. Am. Chem. Soc.* **130** 3296
[2] Chen G F, Li Z, Wu D, Dong J, Li G, Hu W Z, Zheng P, Luo J L and Wang N L 2008 *Phys. Rev. Lett.* **100** 247002
[3] Cheng P, Fang L, Yang H, Zhu X Y, Mu G, Luo H Q, Wang Z S and Wen H H 2008 *Sci. China Ser. G* **51** 719
[4] Ren Z A, Che G C, Dong X L, Yang J, Lu W, Yi W, Shen X L, Li Z C, Sun L L, Zhou F and Zhao Z X *Europhys. Lett.* 2008 **83** 17002
[5] Rotter M, Tegel M and Johrendt D 2008 *Phys. Rev. Lett.* **101** 107006
[6] Wang X C, Liu Q Q, Lv Y X, Gao W B, Yang L X, Yu R C, Li F Y and Jin C Q 2008 *Solid State Commun.* **148** 538
[7] Hsu F C, Luo J Y, Yeh K W, Chen T K, Huang T W, Wu P M, Lee Y C, Huang Y L, Chu Y Y, Yan D C and Wu M K 2008 *Proc. Natl. Acad. Sci.* **105** 14262
[8] Ogino H, Katsura Y, Horii S, Kishio K and Shimoyama J 2009 *Supercond. Sci. Technol.* **22** 085001
[9] Medvedev S, McQueen T M, Troyan I A, Palasyuk T, Eremets M I, Cava R J, Naghavi1 S, Casper1 F, Ksenofontov1 V, Wortmann G and Felser C 2009 *Nat. Mater.* **8** 630
[10] Millican J N, Phelan D, Thomas E L, Leão J B and Carpenter E 2009 *Solid State Commun.* **149** 707
[11] Guo J G, Jin S F, Wang G, Wang S C, Zhu K X, Zhou T T, He M and Chen X L 2010 *Phys. Rev.* B **82** 10520(R)
[12] Krzton-Maziopa A, Shermadini Z, Pomjakushina E, Pomjakushin V, Bendele M, Amato A, Khasanov R, Luetkens H, Conder K 2010 *arXiv*:1012.3637
[13] Wang A F, Ying J J, Yan Y J, Liu R H, Luo X G, Li Z Y, Wang X F, Zhang M, Ye G J, Cheng P, Xiang Z J and Chen X H 2010 *arXiv*:1012.5525
[14] Yeh K W, Ke C T, Huang T W, Chen T K, Huang Y L, Wu P M and Wu M K 2009 *Crystal Grwoth & Design.* **9** 4847
[15] Yadav C S and Paulose P L 2009 *New J. Phys.* **11** 103046
[16] Kida T, Matsunaga T, Hagiwara M, Mizuguchi Y, Takano Y and Kindo K 2009 *J. Phys. Soc. Jpn.* **78** 113701
[17] Braithwate D, Lapertot G, Knafo W and Sheikin I 2010 *J. Phys. Soc. Jpn.* **79** 053703
[18] Taen T, Nakajima Y and Tamegai T 2010 *Physica* C **470** S391
[19] Taen T, Tsuchiya Y, Nakajima Y and Tamegai T 2009 *Phys. Rev.* B **80** 092502
[20] Taen T, Nakajima Y and Tamegai T 2010 *Physica* C **470** 1106
[21] Mizuguchi Y, Deguchi K, Tsuda S, Yamaguchi T, Takeya H, Kumakura H and Takano Y 2009 *Appl. Phys. Express* **2** 083004
[22] Ozaki T, Deguchi K, Mizuguchi Y, Kumakura H and Takano Y 2010 arXiv:1008.1447
[23] Ozaki T, Deguchi K, Mizuguchi Y, Kawasaki Y, Tanaka T, Yamaguchi T, Tsuda S, Kumakura H and Takano Y 2011 *arXiv*:1103.0402
[24] Ding Q, Taen T, Mohan S, Nakajima Y and Tamegai T 2011 *Physica* C accepted
[25] Soibel A, Zeldov E, Rappaport M, Myasoedov Y, Tamegai T, Ooi O, Konczykowski M and

Geshkenbein V B 2000 *Nature* **406** 282
[26] Yasugaki M, Itaka K, Tokunaga M, Kameda N, Tamegai T 2002 *Phys. Rev*. B **65** 212502
[27] Sales B C, Sefat A S, McGuire M A, Jin R Y and Mandrus D 2009 *Phys. Rev*. B **79** 094521
[28] Yeh K W, Huang T W, Huang Y L, Chen T K, Hsu F C, Wu P M, Lee Y C, Chu Y Y, Chen C L, Luo J Y, Yan D C and Wu M K 208 *Europhys. Lett*. **84** 37002
[29] Mizuguchi Y, Tomioka F, Tsuda S, Yamaguchi T and Takano Y 2009 *J. Phys. Soc. Jpn*. **78** 074712
[30] Awana S, Pal A, Vajpayee A, Mudgel M, Kishan H, Husain M, Zeng R, Yu S, Guo Y F, Shi Y G, Yamaura K, Takayama-Muromachi E 2010 *J. Appl. Phys*. **107** 09E128
[31`] Liu T J, Ke X, Qian B, Hu J, Fobes D, Vehstedt E K, Pham H, Yang J H, Fang M H, Spinu L, Schiffer P, Liu Y and Mao Z Q 2009 *Phys. Rev*. B, **80** 174509
[32] Werthamer N R, Helfand E and Hohenberg P C 1966 *Phys. Rev*. **147** 295
[33] Gurevich A, Patnaik S, Braccini V, Kim K H, Mielke C, Song X, Cooley L D, Bu S D, Kim D M, Choi J H,Belenky L J, Giencke J, Lee M K, Tian W, Pan X Q, Siri A, Hellstrom E E, Eom C B and. Larbalestier D C 2004 *Supercond. Sci. Technol*. **17** 278
[34] Bean C P 1964 *Rev. Mod. Phys*. **36** 31
[35] Tamegai T, Nakajima Y, Tsuchiya Y, Iyo A, Miyazawa K, Shirage M P, Kito H and Eisaki H 2008 *J. Phys. Soc. Jpn. Suppl*. C **77** 54
[36] Tamegai T, Nakajima Y, Tsuchiya Y, Iyo A, Miyazawa K, Shirage M P, Kito H and Eisaki H 2008 *Physicsa* C **469** 915
[37] Nakajima Y, Maruoka T, Tamegai T, Kamihara Y, Hirano M and Hosono H *Physica* C **470** S406
[38] Wang L, Qi Y P, Wang D L, Gao Z S, Zhang X P, Zhang Z Y, Wang C L and Ma Y W 2010 *Supercond. Sci. Technol*. **23** 075005
[39] Qi Y P, Wang L, Wang D L, Zhang Z Y, Gao Z S, Zhang X P and Ma Y 2010 *Supercond. Sci. Technol*. **23** 055009
[40] Deutscher G 2010 *Appl. Phys. Lett*. **96** 122502
[41] Lee S, Jiang J, Weiss J D, Folkman C M, Bark C W, Tarantini T, Xu A, Abraimov D, Polyanskii A, Nelson C T, Zhang Y, Baek S H, Jang H W, Yamamoto A, Kametani F, Pan X Q, Hellstrom E E, Gurevich A, Eom C B and Larbalestier D C 2009 *Appl. Phys. Lett*. **95** 212505
[42] Li S, Fistl M, Deak J, Mtecalf P and McElfresh M 1995 *Phys. Rev. B* **52** R747

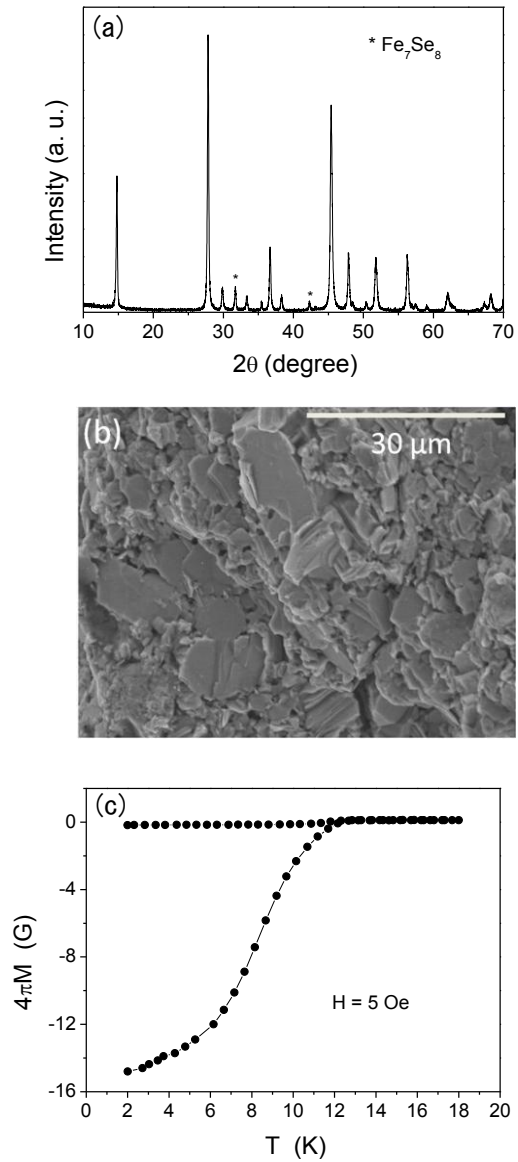

**Figure 1.** (a) Powder X-ray diffraction pattern of as-prepared FeTe$_{0.5}$Se$_{0.5}$ polycrystalline sample. (b) SEM image of the FeTe$_{0.5}$Se$_{0.5}$ polycrystalline sample. (c) Temperature dependence of magnetic susceptibility of FeTe$_{0.5}$Se$_{0.5}$ polycrystalline sample measured at 5 Oe.

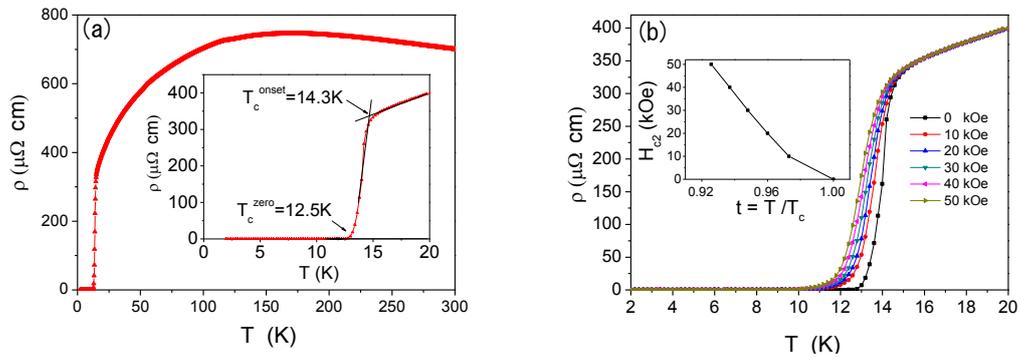

**Figure 2.** (a) Temperature dependence of the resistivity of FeTe$_{0.5}$Se$_{0.5}$ polycrystalline sample. Inset shows the resistivity around $T_c$. (b) Field dependence of the resistivity of FeTe$_{0.5}$Se$_{0.5}$ polycrystalline sample around $T_c$. Inset shows the upper critical field $H_{c2}$ versus temperature determined by the midpoint of the resistive transition.

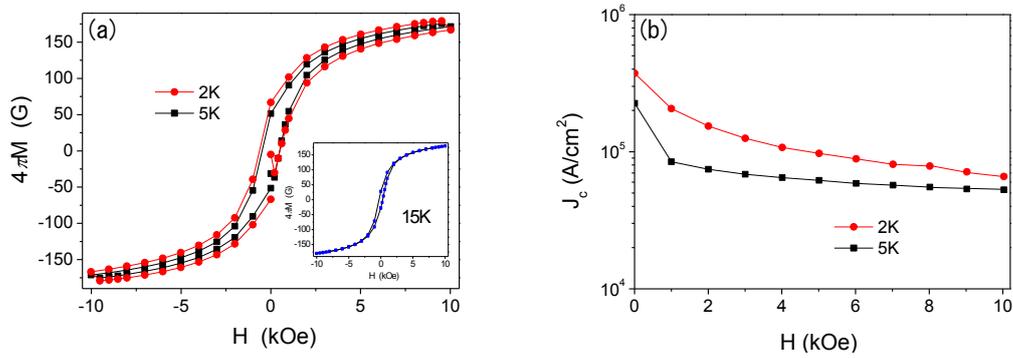

**Figure 3.** (a) Magnetic field dependence of magnetization at different temperatures for FeTe$_{0.5}$Se$_{0.5}$ polycrystalline sample. Inset shows *M-H* curves at 15 K. (b) Magnetic field dependence of critical current densities calculated from the data in (a).

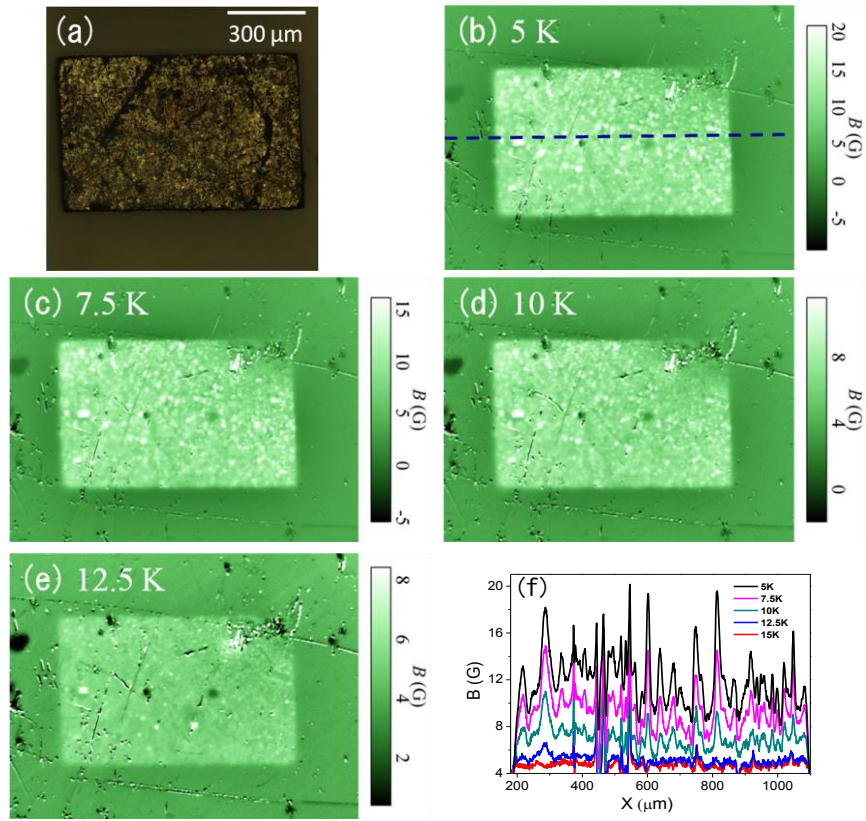

**Figure 4.** (a) Optical images of the FeTe$_{0.5}$Se$_{0.5}$ polycrystalline sample used for MO imaging. Differential MO images in the remanent state of the FeTe$_{0.5}$Se$_{0.5}$ polycrystalline sample at (b) 5 K, (c) 7.5 K, (d) 10 K, and (e) 12.5 K after cycling the field up to 500 Oe for 0.25 seconds, (f) the local magnetic induction profiles at different temperatures taken along the dotted lines in (b).

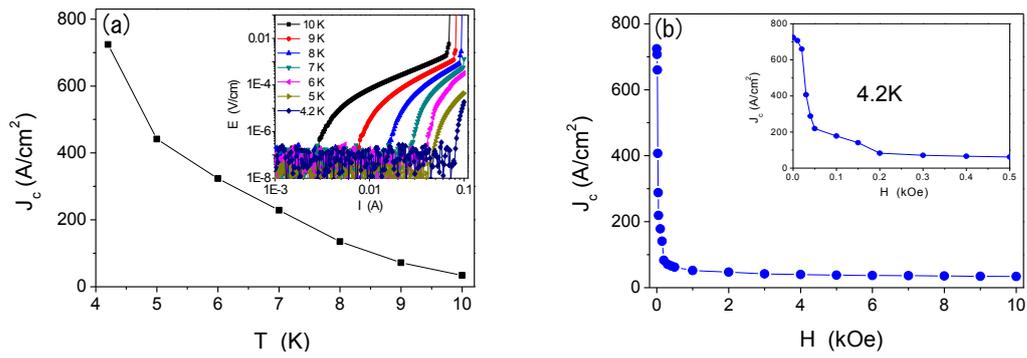

**Figure 5.** (a) Temperature dependence of the transport critical current density $J_c$ in $FeTe_{0.5}Se_{0.5}$ polycrystalline sample at zero field. Inset shows *I–V* characteristics in $FeTe_{0.5}Se_{0.5}$ polycrystalline sampl at *T* = 4.2, 5, 6, 7, 8, 9 and 10 K. (b) Field dependence of the transport $J_c$ in $FeTe_{0.5}Se_{0.5}$ polycrystalline sample at 4.2 K. Inset shows $J_c$ at lower fields.

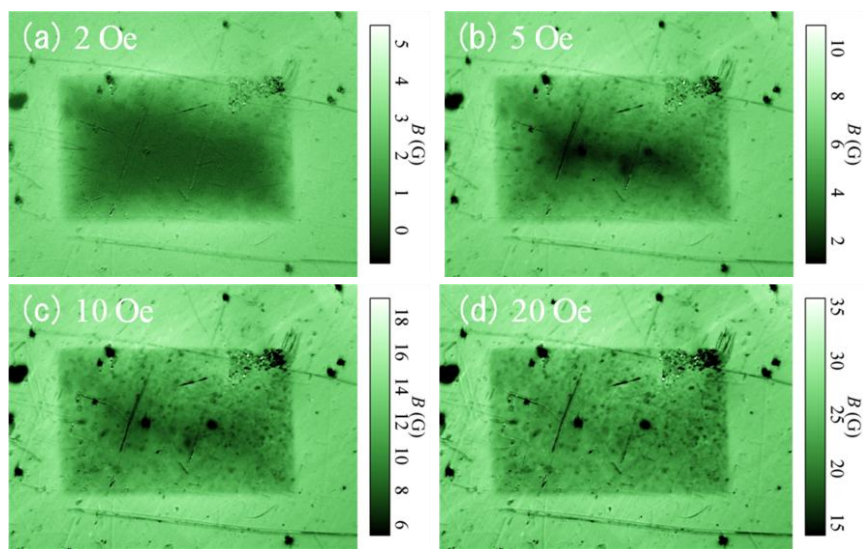

**Figure 6.** MO images of flux penetration into FeTe$_{0.5}$Se$_{0.5}$ polycrystalline sample at (a) 2 Oe, (b) 5Oe, (c) 10 Oe, (d) 20 Oe after zero-field cooling down to 5 K.